\documentclass{article}
\usepackage{graphicx}

\usepackage{epstopdf}

\date{~}

\newcommand{\erf}{\mathop{\rm erf}}
\newcommand{\radon}[1]{\widehat{#1}}
\newcommand{\uvec}[1]{\widehat{\bf #1}}

\begin{document}

\title{%
\vskip-6pt \hfill {\rm\normalsize MCTP-XX-XX} \\ 
\vskip-6pt \hfill {\rm\normalsize  August 2003} \\
\vskip-18pt~\\
Detectability of Weakly Interacting Massive Particles 
in the Sagittarius Dwarf Tidal Stream}
 
\author{Katherine Freese\rlap{,}{$^{1,4}$} Paolo Gondolo\rlap{,}{$^{2,4}$}
  Heidi Jo Newberg\rlap{$^{3}$}
  \\~\\
  \small \it ${}^{1}$Dept.\ of Physics, University of Michigan,
  Ann Arbor, MI 48109\@.  Email: {\tt ktfreese@umich.edu}\@.
  \\
  %\small \it ${}^{2}$Dept.\ of Physics, Case Western Reserve University,
  %\\
  %\small \it 10900 Euclid Ave., Cleveland, OH 44106-7079, USA. Email:
  %{\tt pxg26@po.cwru.edu}\@.
  \small \it ${}^{2}$Dept.\ of Physics, University of Utah, 115 S 1400 E \#201, 
  \\
  \small \it Salt Lake City, UT 84112-0830, USA. Email:
  {\tt paolo@physics.utah.edu}\@.
  \\
  \small \it ${}^{3}$ Rensselaer Polytechnic Institute, Department of Physics,
  Applied Physics, and Astronomy,
  \\
  \small \it 110 8th Street, Troy, NY 12180-3590\@. Email: {\tt heidi@rpi.edu}
  \\
  \small \it ${}^{4}$ Michigan Center for Theoretical Physics, 3444A Randall Laboratory, 
  \\
  \small \it  University of Michigan, 500 E. University Ave., Ann Arbor, MI 48109-1120
  \vspace{-2\baselineskip}
  }

\maketitle
 
\begin{abstract} 
  \noindent
  Tidal streams of the Sagittarius dwarf spheroidal galaxy (Sgr) may
  be showering dark matter onto the solar system and contributing
  $\sim $(0.3--23)\% of the local density of our Galactic Halo.  If 
  the Sagittarius galaxy contains WIMP dark matter, the extra contribution 
  from the stream gives rise to a step-like feature in the energy recoil
  spectrum in direct dark matter detection.  For our best estimate of
  stream velocity (300 km/sec) and direction (the plane containing the
  Sgr dwarf and its debris), the count rate is maximum on June 28 and
  minimum on December 27 (for most recoil energies), and the location
  of the step oscillates yearly with a phase opposite to that of the
  count rate.  
% The energy of the step should be above the threshold of
% the DAMA/NaI detector.  Thus the WIMP signal from the Sgr stream may
% already be present in the DAMA/NaI data (at a level up to
% 100$\sigma$), and may be useful to help establish the interpretation
% of the DAMA annual modulation as due to WIMPs. In addition, the WIMP
% parameters that best fit the data would need to be recalculated.
% The stream may explain the discrepancy between DAMA results and data
% from other experiments.  
  In the CDMS experiment, for 60 GeV WIMPs,
  the location of the step oscillates between 35 and 42 keV, and for
  the most favorable stream density, the stream should be detectable
  at the 11$\sigma$ level in four years of data with 10 keV energy
  bins.  Planned large detectors like XENON, CryoArray and the
  directional detector DRIFT may also be able to identify the Sgr
  stream.
\end{abstract}

\section{Introduction}

The dark halo of our galaxy may consist of WIMPs (weakly interacting
massive particles)\@.  Direct detection experiments attempt to observe
the nuclear recoil caused by these dark matter particles interacting
with the nuclei of the detector.  Currently many such detectors are in
operation worldwide, including CDMS, COSME, CRESST, DAMA, DRIFT,
Edelweiss, Elegant V, HDMS, IGEX, LiF/Tokyo, LosAlamos, MiBeta,
ROSEBUD, Saclay, UKDMC, and Zeplin. These detectors may
be able to observe not only the general halo component but also
streams of dark matter moving through our solar system.

Recently, the Sloan Digital Sky Survey (SDSS) and the Two Micron
All Sky Survey (2MASS) surveys
\cite{newberg,majewski} have traced the tidal stream of the
Sagittarius dwarf galaxy (Sgr) more than $360^\circ$ around the sky.
It is believed that this tidal stream is currently moving through the
solar neighborhood, and that it contains dark matter in addition to
stars.  In this paper we investigate the possibility of detecting WIMP
dark matter in this stream via direct detection experiments.

The Sagittarius galaxy, a dwarf spheroidal galaxy of roughly
$10^9 M_\odot$, is a satellite of our own much larger Milky Way
Galaxy, located inside the Milky Way, $\sim$15 kpc behind the Galactic
Center and $\sim$6 kpc below the Galactic Plane \cite{ibata95}\@.
There are two streams of matter that extend outwards from the main
body of the Sgr galaxy.  These streams, known as the leading and
trailing tidal tails, are made of matter tidally pulled away from the
Sgr galaxy.  The leading tail may be showering matter down
upon the solar system \cite{majewski}\@.  The flow is in the general
direction orthogonal to the Galactic plane and has a speed of roughly
300 km/s (see discussion in \S 2 below)\@. This speed is comparable to
that of the relative speed of the Sun and the WIMPs in the general
dark halo. Hence one can hope to detect the stream in direct detection
experiments.

The detectability depends on the density of dark matter in the stream.
The mass-to-light ratio $M/L$ in the stream is unknown, but is
plausibly at least as large as that in the Sgr main body; in fact, the
$M/L$ in the stream may be significantly larger because the dark
matter on the outskirts of the main body would be tidally stripped
before the (more centrally located) stars.  Various
determinations of the $M/L$ for the Sgr main body give values in the
range 25 to 100 (see the discussion in \cite{majewski} and references
therein)\@.  With our assumptions, we find a density
of dark matter in the stream in the range 0.3 to 23\% of the local dark
halo density.  If the dwarf does not contain dark matter or the stream 
does not pass through the solar neighborhood, then the contribution to 
the local dark matter density is zero.

This additional flux of WIMPs from the stream shows up as a 0.3--23\%
increase in the rate of nuclear recoils at energies below a
characteristic energy $E_c$, the highest energy that WIMPs in the
stream can impart to a target nucleus.  Hence, there is a step in the
energy recoil spectrum; the count rate in the detector is enhanced at
low energies, but then returns to the normal value (due to Galactic
Halo WIMPs) at all energies above the critical energy $E_c$\@.  This
feature can be observed as a sharp decrease in the count rate above a
characteristic energy that depends on the mass of the target nucleus,
the mass of the WIMP, and the speed of the stream relative to the
detector.

It is well known that the count rate in WIMP detectors will experience
an annual modulation \cite{dfs, ffg} as a result of the motion of the
Earth around the Sun: the relative velocity of the detector with
respect to the WIMPs depends on the time of year.  We find that the
count rate of WIMPs in the stream is a maximum on June 28\footnote{We
  suggest that readers prepare their WIMP shields.} and a minimum on
December 27\@.\footnote{For energies near $E_c$, there is a small
  range of energy recoils in which the count signal has the same phase
  as the characteristic energy and so peaks in December.} Note that
this modulation has a different phase than does the ordinary annual
modulation of WIMPs in the Galactic halo, which peaks on June 2\@ for
high energy recoils (and December 20 for low energy thresholds
\cite{hasenbalg,primack,lewinsmith,lf}).
Similarly, we find that the value of the characteristic energy $E_c$
of the step in the energy recoil spectrum oscillates in time with a
period of a year.  The step location experiences an annual modulation
exactly 180$^\circ$ out of phase with that of the stream count rate.
To give a number, for a 60 GeV WIMP detected in a Ge-73 detector, the
characteristic energy for the Sgr stream varies between 25 and 35 keV.
As we will show, this step feature in the energy recoil spectrum
should be detectable at the three $\sigma$ level in the CDMS
experiment using two kilograms of $^{73}$Ge and taking four years of
data with 10 keV energy bins.

%For WIMP masses heavier than 50 GeV, the step lies above the threshold
%energy of the DAMA experiment, and DAMA may have it in their current
%data.  In the current DAMA data, for a 60 GeV WIMP mass and a very
%reasonable stream density that is 4\% of the local halo, the stream is
%detectable at the 24$\sigma$ level in the 2-3 keV electron equivalent
%energy bin.  For the most favorable stream density (20\% of the local
%halo), the stream is detectable at the 104$\sigma$ level in the 2-3
%keV electron equivalent energy bin. In the presence of a stream, the count
%rate peaks on a different date than that of an isothermal halo and for 
%certain combinations of stream densities and stream velocities the date 
%of the peak rate may be in disagreement with current DAMA data. However, 
%with our assumptions on the density and velocity of the Sgr stream,
%the date of the peak rate is within 2$\sigma$ of the current DAMA data. 
% If the existence of
%the Sgr stream were confirmed in the data, then a new best fit WIMP
%mass would have to be obtained.  In addition, the DAMA experiment can
%check for the annual modulation of the step due to the Sgr stream dark
%matter.  Finding the Sgr modulation may be an important step in
%interpreting the observed annual modulation in DAMA in terms of WIMPs.
%It is conceivable that the stream is
%responsible for an apparent discrepancy between DAMA and other
%experiments such as CDMS, EDELWEISS, and ZEPLIN-I.  

In 1994 and again in 2001, the HEAT collaboration \cite{heat1,heat2}
detected an excess of cosmic ray positrons.  Baltz et
al.~\cite{baltz1,baltz2} explored the possibility that this excess is
due to the annihilation of neutralino dark matter in the Galactic
halo. They found that neutralino annihilation can produce enough
positrons to make up the measured excess in the Minimal Supersymmetric
Standard Model provided there is an additional enhancement to the
signal due to dark matter clumps in the Galactic halo\footnote{An
  alternate explanation within the MSSM, namely neutralinos that were
  produced non-thermally in the early Universe, was suggested by Kane
  et al.~\cite{kane1,kane2}.}.  Baltz et al.\ found that, under
current constraints on the MSSM, the ``boost factor'' needs to be at
least 30 in the signal.  The Sagittarius stream provides at most a
23\% enhancement of the signal, and hence is unlikely to provide an
explanation of the boost factor required for neutralino annihilation
to explain the HEAT data.  However, the uncertainties in the
propagation of cosmic rays in the galaxy are large and it may be
possible that a smaller boost factor may suffice, so further studies
are warranted.

Detection of this dark matter stream would be important also for our
understanding of galaxy formation.  The traditional assumption that
the halo population of stars is old and well mixed has been seriously
challenged by recent observations that the outer parts of the stellar
halo contain tidal debris that is merging with the Milky Way even at the 
current epoch.  Recent N-body simulations of galaxy formation suggest that
there may be dwarf satellites of the Milky Way that contain only dark
matter and no stars.
%There is a growing body of
%evidence that at least part of the Milky Way Galaxy was formed through
%the accretion of smaller satellite galaxies, and is not a relic of the
%initial collapse of a single object.  
If dark matter can be detected in the Sagittarius stream, then we could 
use dark matter detectors on Earth to look for remnants of these dark 
companions of our Milky Way.

\section{Local dark matter from the leading tail of Sagittarius}

The factors that affect dark matter direct detection experiments are
the local dark matter density and velocity vector of 
material in the leading tail of the Sgr dwarf galaxy that passes
through the solar neighborhood.
Below, we estimate that the (local) Sgr leading tail is moving
with a velocity of 300$\pm$90 km/s in the direction $0.233 \hat{Y} -
0.970 \hat{Z}$, where $\hat{Z}$ is the north Galactic pole, and
$\hat{Y}$ is in the direction of the local disk stars' motion
($l=90^\circ$)\@.  Equivalently, the local motion of the Sgr leading
tail points in the direction $(l,b)=(90^\circ,-76^\circ)$, where $l$
is the Galactic longitude and $b$ is the Galactic latitude. 
All velocities are measured with respect to the rest frame of the center 
of mass of the Milky Way galaxy.  We will
also find an estimated range for the local density of dark matter of:
\begin{equation}
\rho_{{\rm tail}, \odot} = [2, 190] \times 10^4 
M_\odot/{\rm kpc}^3 = [0.001, 0.07] \, {\rm GeV/cm^3} ,
\end{equation}
which corresponds to (0.3-23)\% of the local
density of the isothermal Galactic halo.

The density calculation is highly uncertain, since (1) the stellar
density is not well measured anywhere along the stream and has not
been directly detected in the solar neighborhood, and (2) $M/L$ is
highly uncertain in the Sagittarius dwarf itself and is unmeasured in
the tidal stream.  Models of the tidal disruption have not been
developed or constrained by observations to the point that they
usefully predict the local properties of coherent dark matter in the
solar neighborhood.  We will derive our estimates from empirical
observations of the old populations of stars in the tidal stream.
Blue horizontal branch stars and main sequence turnoff stars are old
enough that their populations are not likely to have significantly
changed in the time since these stars were stripped from the dwarf
galaxy.  The important measurable quantities are (1) the number of
stars observed in some portion of the tidal stream, (2) the volume of
that portion of the tidal stream, (3) the relationship between number
of observed stars of a particular type and total luminosity in stars,
and (4) $M/L$ in the tidal stream.

\subsection{Velocity in the solar neighborhood}

The Sgr dwarf galaxy is currently about 16 kpc from the Galactic
center, near its closest approach to the Galactic center
(perigalacticon)\@.  As seen from the Sun, Sgr has a radial velocity of 
$140\pm2$ km/s and a tangential velocity of $250 \pm 90$ km/s 
\cite{ibata97}\@.  Correcting for the solar motion as given in 
\S 3.2, we derive a current velocity of $300 \pm 90$ km/s for the 
dwarf, in the reference frame of the Galaxy center.

Stars in the leading tidal tail, which extends to near the solar position, 
were stripped from the side of the dwarf that is closer to the center of 
the Milky Way.  At the time that the stars
are stripped from the gravitational pull of the dwarf galaxy and released
into the potential of the Milky Way, they have about the same orbital 
velocity as the dwarf galaxy itself (we assume that the dwarf galaxy stars
have no bulk rotation).  Due to their closer proximity to the
Galactic center, they have a lower gravitational potential energy, and
therefore shorter orbital periods - thus creating the leading tail.

Stars in the trailing tail were stripped at larger distances from the 
center of the Milky Way.  The higher orbital energy of these stars produces
larger distances from the Galactic center and longer orbital periods.

The trailing tidal tail has been traced in Two Micron All Sky Survey
(2MASS; \cite{2mass}) M giant
stars through nearly $180^\circ$ of sky south of the Galactic plane,
to a maximum distance of $\sim$50 kpc from the Galactic center.  The tidal
tails lie on a plane that passes through the Galactic center.  The Sun
lies within a kiloparsec of that plane, and near the path of the leading
Sgr debris \cite{majewski}.

The stars in the tidal tails must conserve both 
energy and angular momentum, in the absence of dynamical friction.
Since the stars in the trailing tail of Sgr have an apogalacticon of 90 kpc 
or more, while the stars in the leading tail have an apogalacticon of only
about 45 kpc \cite{newberg}, the trailing tail stars must have 
significantly higher energies than the stars in the leading tail.  
If most of the stars are
stripped near the perigalacticon of the Sagittarius dwarf, at 16 kpc
from the center of the Milky Way, then the leading and trailing debris 
must have been stripped from parts of the Sagittarius dwarf that differ in
Galactocentric distance by at least 10 kpc \cite{lapalma}.  It is sensible,
therefore, that the perigalacticon of the stars in the leading tail could 
be 11 kpc or less.

Since the mass of the Milky Way is distributed throughout its volume and
is not concentrated at the center, stars do not travel in closed orbits, but
rather in rosette patterns.  The angle between successive perigalacticon 
passages is less than 360$^\circ$.  The Sun is near the first expected
perigalacticon passage of the leading tail, at about the right distance.
Since the tidal stream is approximately 6 kpc across, it is reasonable
to suspect that the tidal steam passes through the solar position.
Majewski et al. \cite{majewski} show that stellar debris appears to be raining
down on the solar position from approximately the direction of the north
Galactic pole.

Without significant dissipative forces, the average speed of the tidal
debris in the solar neighborhood (which is at perigalacticon) should
be the same as the current speed of the Sgr dwarf (which is also at
perigalacticon).  The debris should travel approximately in the plane
of the Sgr dwarf tidal debris.  If the debris is near perigalacticon,
then its motion is approximately perpendicular to the line from the
Sun to the Galactic center.  From these three conditions, we derive that
the debris velocity is in the direction $0.233 \hat{Y} - 0.970 \hat{Z}$,
where $\hat{Z}$ is the north Galactic pole, and $\hat{Y}$ is in the
direction of the local disk stars' motion ($l=90^\circ$)\@. In terms of 
galactic latitude $b$ and galactic longitude $l$, this direction is 
equivalent to the direction $(l,b)=(90^\circ,-76^\circ)$.

\subsection{Simple estimate of local dark matter density}

In this section we use simple parameters of the Sgr dwarf tidal system
to roughly estimate the Sgr dwarf contribution to the local dark matter
density.  In the next section we will make more careful estimates of 
the parameters.

Published estimates of the original total mass of the Sagittarius dwarf galaxy
are in the range $10^9$ to $10^{11} M_\odot$\@.  The lower number is
within a factor of two of current estimates of the mass of the Sgr
dwarf, and the larger number is derived from the need to keep the
dwarf intact for a Hubble time \cite{jiangbinney2000}\@.  The width of
the tidal stream in 2MASS M stars is estimated to be 4-8 kpc
\cite{majewski}\@.  Below, we will calculate a FWHM of 6 kpc from the
data from the SDSS \cite{newberg2002}\@.  This width is of the order of
the size of the dwarf galaxy.

%A low estimate for the local density would assume the dwarf started
%out with only $10^9 M_\odot$, that half of the mass is still in the
%dwarf and half in the tail, and that the tidally stripped dark matter
%is scattered through a cylinder that extends 10 kpc above and 10 kpc
%below the plane of the Sagittarius dwarf orbit and has a radius of 100
%kpc.  This contains the entire volume in which Sgr tidal debris has
%been identified.  Then, the local dark matter density is:
%\begin{equation}
%\rho_{{\rm tail},\odot} = \frac{5 \times 10^8 M_\odot}{\pi (100 {\rm ~kpc})^2 (20 {\rm ~kpc})} = 800 M_\odot/{\rm kpc}^3 .
%\end{equation}
%
%An upper limit can be estimated by assuming the dwarf started out with
%$10^{11} M_\odot$, and that dark matter is spread evenly along the
%tidal stream.  The tidal stream has been traced over a length of 250
%kpc and is at least 4 kpc in width.
%\begin{equation}
%\rho_{{\rm tail},\odot} = \frac{10^{11} M_\odot}{(4 {\rm ~kpc})^2 (250 {\rm ~kpc})} = 2.5 \times 10^7 M_\odot/{\rm kpc}^3 .
%\end{equation}

Estimates of the {\it current} mass of the Sgr dwarf galaxy are in the range
\begin{equation}
\label{eq:sgrmass}
M_{Sgr} \sim [5,20] \times 10^8 M_\odot ,
\end{equation}
possibly several orders of magnitude lower than the original mass of
the Sgr dwarf at its time of formation.  The lower mass is the mass
within the Sgr dwarf tidal radius calculated by Majewski et al.
\cite{majewski} from the dispersion of the radial velocities of Sgr
dwarf stars.  Comparing their mass to the luminosity of the Sgr, they
derive a mass-to-light ratio $M/L$ of 25-29 in solar units.  Other
authors have argued for $M/L$ as high as 100 \cite{ibata97} (four
times as high) based on the need to keep the Sgr dwarf intact for a
Hubble time.  This produces the upper end of the current mass range.
Given that the $M/L$ ratio is expected to be 25 or higher, calculations of
the mass density in the Sgr dwarf and tidal tails are essentially
equal to the dark matter density.

A typical disruption model \cite{johnston} assumes the current mass
of the Sgr dwarf is about $10^9 M_\odot$, and that the dwarf lost about 2-3
times that mass in the last gigayear (somewhat longer than an orbit).
If this is spread out into a tidal stream 6 kpc in diameter and 250
kpc in length (this is the length of the observed tidal stream in stars),
then an estimate of the local dark matter density is:
\begin{equation}
\rho_{{\rm tail},\odot} = \frac{2.5 \times 10^{9} M_\odot}{(6 {\rm ~kpc})^2 (250 {\rm ~kpc})} = 3 \times 10^5 M_\odot/{\rm kpc}^3 .
\end{equation}
This is one percent of the estimated local dark matter density.

\subsection{Detailed calculation of the stream density}

We now explore the range of dark matter local dark matter densities that are
consistent with our knowledge of the Sagittarius system, assuming that the
leading tidal tail passes through the solar neighborhood.  

In order to calculate the local dark matter density from data, we
would like to measure the local mass density of stars from the stream,
and multiply by $M/L$ calculated for our position on the leading tidal
tail.  Neither the local density of Sgr stream stars, nor the local
mass-to-light ratio in the stream, is known.  Instead, we can
calculate the density of F/G turnoff stars at one location along the
trailing tidal tail, tabulate the blue horizontal branch star
density at several positions along the leading and trailing tidal
streams, and estimate $M/L$ from the Sgr dwarf itself.  Using these
quantities, we estimate the dark matter density from the Sgr tidal
stream in the solar neighborhood.

Specifically, we wish to estimate $\rho_{{\rm tail}, \odot}$, the
dark matter density of the tail in the neighborhood of the Sun, as the 
product of three quantities:
\begin{equation}
\label{eq:product}
\rho_{{\rm tail}, \odot} = \left(\frac{\rm number~of~stars}{\rm volume}\right)
\left(\frac{\rm luminosity}{\rm number~of~stars} \right)
\left(\frac{\rm mass}{\rm luminosity} \right).
\end{equation}
Ideally, one would like to measure all of these quantities in the
tidal stream at the solar position.  Because none of these quantities
are known at the solar position, each term will be estimated from
other measurable portions of the Sgr tidal stream.

To estimate the first term on the right hand side, we will measure the density
of F/G turnoff stars in a portion of the trailing tidal tail 50 kpc
from the Sgr dwarf (stripe 82 in the SDSS).  The density in this location
is the best measured anywhere in the tidal tails.  To obtain an estimate of the
density at our location, we extrapolate from measurements of Blue Horizontal 
Branch (BHB) stars at three places along the tidal tails.  We discover
that the density of the stars in a piece of the leading tail 50 kpc
from the Sgr dwarf (stripe 10 in the SDSS) is nearly a factor of two
higher, giving us a larger range of possible dark matter densities at
the solar position: $\rho_{F/G,\odot} \sim \rho_{{\rm F/G},82}
\bigl({\rho_{{\rm BHB},\odot} \over \rho_{{\rm BHB},82}}\bigr)$.  The
second term on the right hand side of Eq.(\ref{eq:product}) is then
${L_{{\rm Sgr}} \over N_{\rm F/G,Sgr}}$.  The third term is
$(M/L)_{\rm tail}$.  Multiplying together these three factors, we find
\begin{equation}
\rho_{{\rm tail}, \odot} \approx
\rho_{{\rm F/G}, 82}
\left(\frac{\rho_{{\rm BHB}, \odot}}{\rho_{{\rm BHB}, 82}}\right)
\left(\frac{M_{\rm Sgr}}{N_{\rm F/G, Sgr}}\right)
\left(\frac{M/L_{\rm tail}}{M/L_{\rm Sgr}}\right)
\end{equation}
where
\begin{eqnarray*}
\rho_{{\rm F/G}, 82} & = & \parbox[t]{3.5in}{the number of Sgr F/G stars per kpc$^3$
  in stripe 82,} \\
\rho_{{\rm BHB}, \odot} & = & \parbox[t]{3.5in}{the density of Sgr BHB stars
  at the solar position,} \\
\rho_{{\rm BHB}, 82} & = & \parbox[t]{3.5in}{the density of Sgr BHB stars
  in stripe 82,} \\
M_{\rm Sgr} & = & \parbox[t]{3.5in}{the current mass of the Sagittarius dwarf,
  } \\
N_{\rm F/G, Sgr} & = & \parbox[t]{3.5in}{the number of F/G stars in
  the Sagittarius dwarf,} \\
M/L_{\rm tail} & = & \parbox[t]{3.5in}{the mass-to-light ratio in the tidal tail,
  } \\
M/L_{\rm Sgr} & = & \parbox[t]{3.5in}{the mass-to-light ratio in the Sgr dwarf.}
\end{eqnarray*}
Each of these terms is discussed in further detail in the following
sections.

\subsubsection{Estimating $\rho_{{\rm F/G}, 82}$}

First, we measure the stellar density of F/G turnoff stars in stripe 82,
$(l,b) = (167, -54)$, about 50 kpc from the Sgr dwarf along the trailing
tidal tail.  Figure 22 of Newberg et
al. \cite{newberg2002} shows the counts of F/G turnoff stars along the
Celestial Equator.  From that figure, one can estimate that in the Sgr
stream overdensity labeled S167-54-21.5 (in red), there are 1000 stars
per half-degree-wide bin at the peak.  It is difficult to estimate the
background star counts in that region of the sky, but it is between
zero and 400 stars per half degree bin.  The total width of the
(approximately triangular) profile is 50 degrees, or 100 bins.  Thus,
we estimate the total number of F/G Sgr stream stars in this 2.5
degree-wide stripe is between 30,000 and 50,000.

In order to calculate the stream density at this position, we need to
know the volume represented by the piece of the stream we have
sampled.  The FWHM of the stream is about $25^\circ$ in the observed
stripe of sky.  At a distance of 29 kpc from the Sun, this would imply
a stream width of 13 kpc.  [Note, however, that the Celestial Equator
does not cut through the Sagittarius stream perpendicular to its path
across the sky.  The angle between the stream direction and the
Celestial Equator is $28^\circ$ \cite{newberg2002}, thus implying that
the actual width of the stream is (13 kpc) $\sin(28^\circ) = 6$ kpc.]

We estimate that the angle between the actual tidal stream and the
projection of the tidal stream perpendicular to our line of sight is
about $10^\circ$ at this position in the sky \cite{newberg,majewski},
so the distance along the stream in the plane of the Celestial Equator
is $(13 {\rm ~kpc})/\cos(10^\circ) = 13$ kpc.  In the $2.5^\circ$ wide
stripe we sample 1.3 kpc perpendicular to the Celestial Equator.  If
the stream is in the range [4,8] kpc deep, then the volume is:
\begin{equation}
V = (13 {\rm ~kpc}) (1.3 {\rm ~kpc}) ([4,8] {\rm kpc}) = [68,140] {\rm kpc}^3.
\end{equation}
Therefore, the density of F/G turnoff stars in that portion of the
stream is:
\begin{equation}
\rho_{{\rm F/G}, 82} = \frac{[30,000; 50,000] {\rm stars}}{[68,140] {\rm kpc}^3} = [210, 740] {\rm stars/kpc}^3.
\end{equation}

\subsubsection{Estimating ${\rho_{{\rm BHB}, \odot}}/{\rho_{{\rm BHB}, 82}}$}

One now asks how much the stellar density varies as a function of
position along the stream.  This is not well known, but we will
estimate it from BHB star counts
\cite{newberg}\@.  It would be better if we could measure the number of
F/G turnoff stars, as they a more stable indicator of stellar density.
However, the only piece of the stream close enough to the Sun that
this quantity has been measured reliably is the stripe 82 data
discussed above.  Table I summarizes the data and calculations
described below.

\begin{table}[tbp]
\begin{tabular}{|l|c|c|c|c|c|l|}
\hline
stripe & A stars & D (kpc) & $\cos(\theta)$ & $\sin(\phi)$ & (A stars) $\times$ & Distance along \\
 &  &  &  &  & $\cos(\theta)\sin(\phi)/D$ & tail from Sgr dwarf\\
\cline{2-7}
\hline
10 & 125 & 48 & 0.8660 & 0.5071 & 1.14 & 50 kpc, leading \\
\hline
29 & 58 & 83 & 0.9848 & 0.9699 & 0.67 & 150 kpc, trailing \\
\hline
82 & 35 & 29 & 0.9848 & 0.5071 & 0.60 & 50 kpc, trailing \\
\hline
\end{tabular}
\caption{Calculation of relative A star density at three places along
the Sgr dwarf tidal tail.  Stripes 29 and 82 cross the trailing tidal
stream, and stripe 10 crosses the leading tidal tail.  Stripes 29 and
10 are near perigalacticon for the trailing and leading tidal tails,
respectively.  The last column gives the adjusted relative A star
density for each piece.}
\end{table}

The stripe number (column 1) refers to the piece of sky scanned in the
SDSS.  Column two gives an estimate of the number of BHBs in the
Sagittarius stream in each of those pieces of data \cite{newberg}\@.
The third column gives the distance from the Sun to the center of the stream,
assuming the absolute magnitude for the BHB stars is $M_g = 0.7$\@.

We define $\theta$ to be the angle between the normal to our line of
sight towards a particular piece of the stream and the tangent to the
stream at that point.  We estimate $\theta = 30^\circ, 10^\circ,
10^\circ$ for stripes 10, 29, 82, respectively \cite{newberg,majewski}\@.  
If $\theta=0^\circ$, then the stream is perpendicular to
our line of sight.  If $\theta$ is larger than that, then we see a
larger distance along the stream in each solid angle of sky.  The
volume within the stream is inversely proportional to $\cos(\theta)$\@.
$\cos(\theta)$ is tabulated in column 4.

If we define $\phi$ as the angle between the observations and the
Sagittarius dwarf stream, then $\cos(\phi)$ is found by dotting the
Sagittarius dwarf plane normal vector with the SDSS observational
plane normal vector.  The normal to the plane of the Sagittarius dwarf
tidal stream is: $\hat{n} = -0.064 \hat{X} + 0.970 \hat{Y} + 0.233
\hat{Z}$ \cite{majewski}\@.  In contrast to the referenced paper, we
have defined $\hat{X}$ to be positive towards the Galactic center from
the Sun.  The stripes 10 and 82 are both on the Celestial Equator, so
the normal to that observational plane is the north Celestial Pole:
$\hat{n}_{pole} = -0.4834 \hat{X} + 0.7472 \hat{Y} + 0.4560 \hat{Z}$\@.
The normal to the plane of observation of stripe 29 is: $\hat{n}_{29}
= -0.1309 \hat{X} + 0.4534 \hat{Y} - 0.8816 \hat{Z}$\@.  If 
$\phi=90^\circ$, then we scan directly across the tidal stream.  Smaller 
angles will net a larger effective volume and more stars; the volume 
is inversely proportional to $\sin(\phi)$\@.  
We tabulate $\sin(\phi) = \sqrt{1-\cos^2(\phi)}$ in column 5 of Table I.

The column 6 in the table above gives a relative density of A stars in
each of the three positions along the stream, corrected for projection
effects ($\theta$), and the angle between the stream and the direction
we scan across it ($\phi$)\@.  Following the calculation of F/G star density
above, the volume sampled in each stripe, for a stream of cross sectional
area $A$, is given by:
\begin{equation}
V = \frac{A D \tan(2.5^\circ)}{\cos(\theta) \sin{\phi}}.
\end{equation}
To estimate A star densities for a stream of diameter 6 kpc, divide the numbers
in column 6 by $\pi (3 {\rm ~kpc})^2 \tan(2.5^\circ) = 1.23 {\rm ~kpc}^{3}$.

Comparing the numbers in column 6, we see that the density in the
trailing tail is similar in the two places sampled.  The density in
the leading tail in the one place sampled is twice as high.  The
Sagittarius tidal debris at the position of the Sun is further along
the leading tail, and we guess from these measurements that it could
have a mass density similar to, or twice as large as, that measured 
in ``stripe 82.''  
\begin{equation}
{\rho_{{\rm BHB}, \odot}}/{\rho_{{\rm BHB}, 82}} = [1,2].
\end{equation}

\subsubsection{Estimating ${M_{\rm Sgr}}/{N_{\rm F/G, Sgr}}$, $M/L$}

All that is left now is to find the conversion from the density of turnoff stars to a
mass density of dark matter.  We will do this by comparing the number
of F/G stars in the Sgr dwarf galaxy to the estimated mass of the Sgr
dwarf.  Figure 6 of Newberg et al. \cite{newberg2002} shows a
color-magnitude diagram of stars from two small fields near the center
of the Sgr dwarf \cite{marconi1998}, shifted into SDSS colors.  The
F/G stars in the stripe 82 data were selected with $0.0 < g-r < 0.6$
and $20.5 < g < 22.5$\@.  This color-magnitude selection was chosen to
maximize the overdensity of Sagittarius stars in stripe 82.  Since the
F/G turnoff stars in Figure 6 of Newberg et al. \cite{newberg2002} are
a magnitude fainter in $g$, we selected all stars in that figure with
$0.0 < g-r < 0.6$ and $21.5 < g < 23.5$\@.  There are 5786 F/G turnoff
stars within these limits.

The field centers of the two Sgr dwarf fields \cite{marconi1998} from
which the data were taken were both within 1.5 degrees of the center
of the Sagittarius dwarf.  Figure 5 of \cite{majewski} shows the
density of M giant stars along the major axis of the Sagittarius
dwarf.  Within 1.5 degrees of the center of the galaxy the density is
reasonably flat.  The surface brightness in $V$ at the center of the
Sagittarius dwarf is $25.2 \pm 0.3$ magnitudes per square arcsecond
\cite{mateo1995,mateo1998}\@.  The integrated apparent magnitude of the
undisrupted portion of the Sgr dwarf is $V=3.63$ \cite{majewski}\@.
Using a distance modulus of 16.9 to the dwarf, the total luminosity is
$1.7 \times 10^7 L_\odot$, comparable to previous estimates
\cite{ibata95, mateo1996}\@.  
This means in one square arcsecond near the center of Sgr, the
brightness is $4.25 \times 10^8$ times less than the brightness of the
entire dwarf galaxy.  So, if we found 5786 F/G turnoff stars in two
fields $8.5' \times 9'$ in size, then the total number of F/G turnoff
stars in the Sagittarius dwarf is:
\begin{equation}
N_{\rm F/G, Sgr} = \frac{5786 \times 4.25 \times 10^8}{2 \times 9' 
\times 8' \times (60 ''/')^2} = 4.7 \times 10^6 {\rm ~F/G ~stars.}
\end{equation}

%As discussed in Eq.(\ref{eq:sgrmass}) in \S 2.2, a reasonable mass
%estimate for the current Sgr dwarf is:
%\begin{equation}
%M_{Sgr} = [5, 20] \times 10^8 M_\odot.
%\end{equation}

The mass-to-light ratio in the Sgr tidal stream has not been measured by any
technique.  In analogy to other galaxies, we expect the luminous matter in
the progenitor Sgr galaxy to be more concentrated than the dark matter.
In tidal disruption, the outer portions of the dwarf galaxy are tidally
stripped before the inner portions; the tidal radius decreases as the mass
of the galaxy decreases.  Therefore, one imagines that $M/L$
in tidal streams is higher than $M/L$ in the originating dwarf, and that the
mass-to-light ratio increases with distance from the progenitor, as the more
distant pieces were the first to be stripped.

Figure 2 of Padmanabhan et al. \cite{padmanabhan}, shows the radial profiles
of luminous matter and dark matter for a stellar Hernquist profile
embedded in an adiabatically compressed dark matter halo.  $M/L$ at
the half light radius is twice the integrated $M/L$ interior to that
radius.  At 1.5 times the half light radius, $M/L$ rises to three
times $M/L$ interior to the half light radius.  At three times the
half-light radius we virtually run out of luminous matter, so $M/L$
becomes very large.

Since our estimates of the luminous matter in the stream are of the same
order of magnitude as the Sgr dwarf itself, the debris at the solar position
probably was not stripped very far from the half light radius.  We adopt
as a reasonable upper limit that $M/L$ in the stream could be a factor
of three higher than the current $M/L$ of the dwarf.
\begin{equation}
\frac{M/L_{\rm tail}}{M/L_{\rm Sgr}} = [1,3].
\end{equation}

%Multiplying the local luminosity density
%by a mass-to-light ratio in the range $[25, 300]$, 
Putting this all back into Eq.~\ref{eq:product}, we derive an
estimated range for the local density of dark matter of:
\begin{eqnarray}
\rho_{{\rm tail}, \odot} & =  &[210,740] \times [1,2] \times \frac{[5,20] \times 10^8}{4.7 \times 10^6} \times [1,3] \nonumber\\
& = & [2, 190] \times 10^4 M_\odot/{\rm kpc}^3 \nonumber\\
& = & [0.001, 0.07] \, {\rm GeV/cm^3}.
\end{eqnarray}
Comparing with the local density of dark matter not in the Sgr stream,
$\rho_h = 0.3$ GeV/cm$^3$, we see that the Sgr stream contributes an
additional 0.3\%--23\% to the total local dark matter density.

Previously Stiff, Widrow, and Frieman \cite{swf} studied the tidal
disruption of satellite galaxies falling into the halo of our own.
They found that, with probability 0(1), the Sun should be situated
within a stream of density $\sim 4$\% of the local Galactic halo
density.  Such a prediction is close to the density
estimated for the Sgr stream if intermediate values for all variables
are used.  

\section{Direct Detection of WIMPs in the Stream}

\subsection{Count Rates}

More than twenty collaborations worldwide are presently developing
detectors designed to search for WIMPs. Although the experiments
employ a variety of different methods, the basic idea underlying WIMP
direct detection is straightforward: the experiments seek to measure
the energy deposited when a WIMP interacts with a nucleus in the
detector \cite{goodmanwitten}.  

If a WIMP of mass $m$ scatters elastically from a nucleus of mass $M$,
it will deposit a recoil energy $E = (\mu^2v^2/M)(1-\cos\theta)$,
where $\mu \equiv m M/ (m + M)$ is the reduced mass, $v$ is the speed
of the WIMP relative to the nucleus, and $\theta$ is the scattering
angle in the center of mass frame. In this paper we consider the
contribution to the detection rate from two dark matter components:
the usual WIMPs in the Halo of our Galaxy plus the additional
contribution from WIMPs in the Sagittarius tidal stream.  We compute
(following, {\it e.g.}, \cite{lesarcs,gondologelmini}) the
differential detection rate per unit detector mass ({\it i.e.},
counts/day/kg detector/keV recoil energy) as
\begin{equation}
\label{eq:rate}
  \frac{dR}{dE} = \frac{ \sigma_0 F^2(q) } { 2 m \mu^2 } 
\bigl[ \rho_h \eta_{h}(E,t) + \rho_{\rm str} \eta_{\rm str}(E,t) \bigr]
\end{equation}
where $\rho_h = 0.3$ GeV/cm$^3$ is the standard local halo WIMP
density (excluding the stream), $\rho_{str} = \rho_{{\rm tail}, \odot}$ is the dark matter
density in the Sgr tidal stream, $\eta_h$ is the mean inverse speed of
WIMPs in the standard Galactic halo (excluding those in the stream),
$\eta_{\rm str}$ is the mean inverse speed of WIMPs in the stream,
$\sigma_0$ is the total nucleus-WIMP interaction cross section, $q =
\sqrt{2 M E}$ is the nucleus recoil momentum, and $F(q)$ is a nuclear
form factor that takes into account the loss of coherence in
WIMP-nucleus interactions for momentum transfers comparable to or
larger than the inverse nuclear radius (we normalize $F(0) = 1$).  For
detectors consisting of different nuclei, like e.g. NaI or CaWO$_4$,
the total rate is a weighted average of the rates for the individual
nuclei with weights equal to the mass fraction of each nucleus in the
detector.  We discuss these quantities in further detail in the
following.

The nuclear form factor $F(q)$ depends on the type of WIMP-nucleus
interactions, namely if they are spin-dependent or spin-independent,
and reflects the mass and spin distributions inside the nucleus.  For
spin-independent interactions, the target nucleus can be approximated
as a sphere of uniform density smoothed by a gaussian \cite{helm}, and the
form factor follows as
\begin{equation}
F(q) = \frac{ 3 [ \sin(qR_1) - qR_1\cos(qR_1) ] } {q^3 R_1^3} \, e^{-q^2s^2/2}
\end{equation}
where $R_1 = ( R^2 - 5 s^2)^{1/2}$, $ s \simeq 1$ fm, and
\begin{equation}
R \simeq \bigl[0.91 (M/{\rm GeV})^{1/3} + 0.3 \bigr]
\times 10^{-13} {\rm cm}
\end{equation}
is the radius of the nucleus.  For spin-dependent interactions, the form
factor is somewhat different but again $F(0)=1$.  In general the form factor
needs to be evaluated for specific detector nuclei.

For purely scalar interactions,
\begin{equation}
\label{eq:scalar}
\sigma_{0,\rm scalar} = {4 \mu^2 \over \pi} [Zf_p + (A-Z)f_n]^2
\,
.
\end{equation}
Here $Z$ is the number of protons, $A-Z$ is the number of
neutrons,
and $f_p$ and $f_n$ are the WIMP couplings to nucleons.
For purely spin-dependent interactions,
\begin{equation}
\sigma_{0,\rm spin} = (32/\pi) G_F^2 \mu^2 \Lambda^2 J(J+1) \, .
\end{equation}
Here $J$ is the total angular momentum of the nucleus and $\Lambda$ is
determined by the expectation value of the spin content of the nucleus (see
\cite{goodmanwitten,dfs,epv,ressell,jkg}).

For the estimates necessary in this paper, we take the WIMP-nucleon
cross section to be $\sigma_p = 7.2 \times 10^{-42}$ cm$^2$, and take
the total WIMP-nucleus cross section to be\footnote{In most instances,
$f_n \sim f_p$ so that the following equation results from
Eq.(\ref{eq:scalar}).} 
\begin{equation}
\sigma_0 = \sigma_p \left(\frac{\mu}{\mu_p}\right)^2 A^2
\end{equation}
where the $\mu_p$ is the proton-WIMP reduced mass, and A is
the atomic mass of the target nucleus.

Information about the WIMP velocity distribution is encoded into the
mean inverse speed $\eta(E,t)$,
\begin{equation}
\label{eq:eta}  
  \eta(E,t) = 
  \int_{u>v_{\rm min}} \frac{f_{\rm d}({\bf u},t)}{u} \, d^3u,
\end{equation}
where 
\begin{equation}
\label{eq:vmin}
v_{\rm min} = \sqrt{\frac{M E}{2\mu^2}}
\end{equation}
 represents the minimum
WIMP velocity that can result in a recoil energy $E$ and
$f_{\rm d}({\bf u},t)$ 
is the (usually time-dependent)
distribution of WIMP velocities ${\bf u}$ relative to the detector.

In this paper, we neglect the rotation of the Earth and take the
reference frame of the detector to be the same as that of the Earth.
Thus we relate the velocity distribution $f_{\rm d}({\bf u},t)$
relative to the detector to the velocity distribution $f({\bf v})$
relative to the Galactic rest frame using
\begin{equation}
 f_{\rm d}({\bf u},t) = f({\bf v}(t)),
\end{equation}
\begin{equation}
 {\bf v}(t) = {\bf u}+{\bf v}_\oplus(t),
\end{equation}
where ${\bf v}_\oplus(t)$ is the velocity of the Earth in the
Galactic rest frame. We discuss the motion of the Earth in the next section.

\subsection{Motion of the Earth in Galactic coordinates}

In this section we write an expression for the velocity of the Earth in the
Galactic rest frame. We use a coordinate system in which $X$ points
toward the Galactic center, $Y$ toward the direction of Galactic
rotation, and $Z$ toward the North Galactic Pole.

For simplicity, we follow \cite{gondologelmini} and neglect the
ellipticity of the Earth orbit and the non-uniform motion of the Sun
in right ascension (an error of less than 2 days in the position of
the modulation maximum and minimum; more precise expressions for the
motion of the Earth can be found in \cite{lewin,green}). Thus we
write the velocity of the Earth in terms of the Sun ecliptic longitude
$\lambda(t)$ as
\begin{equation}
{\bf v}_{\oplus}(t) = {\bf v}_{\rm LSR} + {\bf v}_{\odot} + V_{\oplus} \, 
\left[ \hat{\bf e}_1 \sin \lambda(t) - \hat{\bf e}_2 \cos \lambda(t) \right] .
\end{equation}
Here
\begin{equation}
{\bf v}_{\rm LSR} = (0,220,0) \, {\rm km/s},
\end{equation}
is the velocity of the Local Standard of Rest, which points in the
direction of Galactic rotation, and
\begin{equation}
{\bf v}_{\odot} = (10,13,7) \, {\rm km/s}
\end{equation}
is the Sun's peculiar velocity. (The uncertainty in the Sun's peculiar
velocity is of the order of 0.2 km/s in the $Z$ direction and of as
much as 3 km/s in the $X$ and $Y$ directions \cite{solarmotion}; the
corresponding uncertainty in the phase constant of the modulation is
of the order of several days, larger than the error from neglecting
the ellipticity of the Earth orbit.) Finally,
\begin{equation}
V_{\oplus} = 2 \pi {\rm A.U.}/{\rm yr} = 29.8~{\rm km/s}  .
\end{equation}
is the orbital speed of the Earth.

The unit vectors $\hat{\bf e}_1$ and $\hat{\bf e}_2$ define the plane
of the Earth orbit, the ecliptic. They are in the direction of the Sun
at the spring equinox and at the summer solstice, respectively. In
Galactic coordinates,
\begin{eqnarray}
&&  \hat{\bf e}_1 = ( -0.0670, 0.4927, -0.8676 ) , \\
&&  \hat{\bf e}_2 = ( -0.9931, -0.1170, 0.01032 ) . 
\end{eqnarray}
The Sun ecliptic longitude $\lambda(t)$ can be expressed as a function of 
time $t$ in years with $t=0$ at January 1 as
\begin{equation}
\lambda(t) = 360^\circ \,  (t - 0.218) .
\end{equation}
Here 0.218 is the fraction of year before the spring equinox (March 21).

\subsection{Recoil Spectrum of Galactic and Sgr stream components}

In this paper we consider two different sources of WIMPs that
contribute to count rates in detectors: WIMPs in the Milky Way halo
and WIMPs in the Sgr stream.  The function $\eta(E,t)$ of
Eqs.(\ref{eq:rate},\ref{eq:eta}) is different for these two
contributions.

{\it Galactic Component:} For WIMPs in the Milky Way halo, the most
frequently employed background velocity distribution is that of a
simple isothermal sphere \cite{ffg}.  In such a model, the Galactic
WIMP speeds with respect to the halo obey a Maxwellian distribution
with a velocity dispersion $\sigma_h$ truncated at the escape velocity
$v_{\rm esc}$,
\begin{equation}
  f_h({\bf v}) = \cases{
    \displaystyle \frac{1}{N_{\rm esc} \pi^{3/2} \overline{v}_0^3} 
    \, e^{-{\bf v}^2\!/\overline{v}_0^2} , 
    & for $ |{\bf v}| < v_{\rm esc} $ \cr
    0 , & otherwise. }
\end{equation}
Here 
\begin{equation}
  \overline{v}_0=\sqrt{2/3} \, \sigma_h
\end{equation}
 and 
\begin{equation}
N_{\rm esc} = \erf(z) - 2 z \exp(-z^2) / \pi^{1/2} ,   
\end{equation}
 with $z
= v_{\rm esc}/\overline{v}_0$, is a normalization factor.  For the
sake of illustration, we take $\sigma_h = 270$ km/s and $v_{\rm
  esc} = 650$ km/s.

It can be shown that for the WIMPs in the isothermal halo
\begin{eqnarray}
  \eta_h(E,t) & = &  
  \left\{ 
    \begin{array}{ll}
      \displaystyle
      \frac{1}{2 v_{\oplus}(t) N_{\rm esc}} \left[
        \erf( x+y ) -  \erf( x-y ) -    
        \frac{4}{\sqrt{\pi}} y e^{-z^2} 
      \right], & \hbox{for $x<z-y$}, \\
      \displaystyle
      \frac{1}{2 v_{\oplus}(t) N_{\rm esc}} \left[
        \erf( z ) -  \erf( x-y ) -    
        \frac{4}{\sqrt{\pi}} (z+y-x) e^{-z^2} 
      \right], & \hbox{for $z-y<x<z+y$}, \\
      \displaystyle
      0, & \hbox{for $x>z+y$}, \\
    \end{array}
    \right.
\end{eqnarray}
where $  x = v_{\rm min}/\overline{v}_0$, 
  $ y = v_{\oplus}(t)/\overline{v}_0$, and 
  $ z = v_{\rm esc}/\overline{v}_0 $.

{\it Sgr Stream Component:}
For WIMPs in the Sgr stream, we can at first assume that they all move
at the same velocity 
\begin{equation}
{\bf v}_{\rm str} = (0, 0.233,-0.970) \times 300 {\rm km/s} , 
\end{equation}
and later consider that in addition they have a small velocity
dispersion $\sigma_{\rm str}$ somewhat larger than 20 km/s.  The
additional flux of WIMPs from the stream shows up as a 0.3--23\%
increase in the rate of nuclear recoils below the highest energy that
WIMPs in the stream can impart to a target nucleus.  Hence, there is a
step in the energy recoil spectrum: the count rate in the detector is
enhanced at low energies, but then returns to the normal value (due to
Galactic Halo WIMPs) at all energies above a critical energy $E_c$.
This feature can be observed as a sharp decrease in the count rate at
a characteristic energy that depends on the mass of the target
nucleus, the mass of the WIMP, and the speed of the stream relative to
the detector.

If we neglect the velocity dispersion, the velocity distribution of
the WIMPs in the Sgr stream becomes a Dirac delta function
\begin{equation}
  f_{\rm str}^{(0)}({\bf v}) = 
  \delta\!\left({\bf v} - {\bf v}_{\rm str} \right),
\end{equation}
and their mean inverse speed computed from Eq.~(\ref{eq:eta}) is constant
up to the characteristic energy $E_c(t)$,
\begin{equation}
  \eta_{\rm str}^{(0)}(E,t) =
\frac{\theta(E_{c}(t) -
  E) }{u_{\rm str}(t)} .
\end{equation}
Here $\theta$ is the Heaviside function, and 
\begin{equation}
  u_{\rm str}(t) = | {\bf v}_{\rm str} - {\bf v}_{\oplus}(t) |
\end{equation}
is the relative speed of the WIMPs with respect to the nucleus.
The characteristic energy at
which there is a step in the recoil spectrum is
\begin{equation}
  E_{c}(t) = \frac{ 2 \mu^2 }{M} \left[ u_{\rm str}(t) \right]^2 .
\end{equation}
This characteristic energy is the maximum recoil energy that can be
imparted to the nucleus, and can be obtained as follows: The maximum
momentum transferred from a WIMP to a nucleus occurs when the WIMP
bounces back and is $ q_{\rm max} = 2 \, \mu \, u_{\rm str}(t) $. The maximum
recoil energy of the nucleus then follows as $E_c(t) = q_{\rm
  max}^2/(2M)$.

{\it Effect of Velocity Dispersion of Sgr stream:}
The effect of a velocity dispersion $\sigma_{\rm str}$ in the Sgr
stream is to smooth out the edges of the step.  We assume that the WIMPs in the
Sgr stream follow a Maxwellian velocity distribution with bulk velocity ${\bf
  v}_{\rm str}$,
\begin{equation}
  f_{\rm str}({\bf v}) = \frac{1}{(2\pi\sigma_{\rm str}^2/3)^{3/2}} 
  e^{-3|{\bf v}-{\bf v}_{\rm str}|^2/2\sigma_{\rm str}^2} .
\end{equation}
The mean inverse speed
in this case can be obtained using the expression of $\eta_h(E,t)$
after taking the limit $v_{\rm esc}\to\infty$ and replacing
$v_{\oplus}(t)$ with $u_{\rm str}(t)$ and $\overline{v}_0$ with
$\sqrt{2/3} \, \sigma_{\rm str}$.
\begin{equation}
  \eta_{\rm str}(E,t) = 
    \frac{1}{2 u_{\rm str}(t)} \left[ 
      \erf\left( 
        \frac{ v_{\rm min}+u_{\rm str}(t) }
          { \sqrt{2/3} \, \sigma_{\rm str} } \right) - 
      \erf\left( 
        \frac{ v_{\rm min}-u_{\rm str}(t) }
          { \sqrt{2/3} \, \sigma_{\rm str} } \right)  
    \right] .
\end{equation}

The velocity dispersion of stars in the stream is observed to
be roughly 20-30 km/sec \cite{yanny,majewski1999,dohmpalmer}.  We will here take the velocity dispersion
of the dark matter to lie in the same range, but caution that
this number is very poorly known.

\subsection{Sgr Stream Annual Modulation}

It is well-known in the case of WIMPs in the Galactic Halo that the
signal experiences an annual modulation \cite{dfs, ffg}, with a peak
on June 2 (for an isothermal halo).  Similarly, the WIMPs in the Sgr
stream also have an annual modulation in the count rate.  As we will
show, the count rate of the stream WIMPs peaks in June and is a
minimum in December.  In addition, the position of the step $E_{c}(t)$
depends on time, and so is annually modulated. It is higher in winter
and lower in summer (i.e. its phase is opposite to that of the count
rate).

The time dependence of $E_c$ can be made explicit as follows.
\begin{equation}
  E_c(t) = E_c^{(0)} \left\{ 1 + A_c \cos[ \omega (t - t_c)] \right\} ,
\end{equation}
with $ \omega = 2 \pi/1$~yr, 
\begin{eqnarray}
  E_c^{(0)} & = & \frac{2 \mu^2}{M} \left[
  \left| {\bf v}_{\rm str} - {\bf v}_{\rm LSR} - {\bf v}_\odot \right|^2 
  + V_\oplus^2 \right] ,
\\
  A_c & = & \frac{2 V_\oplus \sqrt{a_1^2+a_2^2}}
 { \left| {\bf v}_{\rm str} - {\bf v}_{\rm LSR} - {\bf v}_\odot \right|^2+V_\oplus^2} ,
\end{eqnarray}
and $t_c$ is the solution of 
\begin{eqnarray}
  \cos \omega t_c = \frac{a_2}{\sqrt{a_1^2+a_2^2}}, \quad \sin \omega t_c = - \frac{a_1}{\sqrt{a_1^2+a_2^2}} .
\end{eqnarray}
Here
\begin{eqnarray}
  a_i \equiv  \hat{\bf e}_i  \cdot ( {\bf v}_{\rm str} - {\bf v}_{\rm LSR} - {\bf v}_\odot ) .
\end{eqnarray}
For our reference case of $v_{\rm str}=300$ km/s in the direction $0.233 \hat{X} - 0.970 \hat{Z}$, 
we find that
the amplitude of the modulation $A_c= 9.2\%$ and that $E_c(t)$ is
maximum on December 27 ($t_c = 0.991$~yr) and minimum about six months later
on June 28 ($t_c = 0.491$~yr).  Other values of $v_{\rm str}$ between
100 km/s and 300 km/s, and other angles within $30^\circ$ of the 
direction downward of the Galactic plane, give a modulation amplitude $A_c$ that can be as low as
2\% or as high as 18\%. The amplitude and phase of the $E_c$
modulation do not depend on the target nucleus, the WIMP mass, or the
density of WIMPs in the stream. They depend on the magnitude and
direction of the Sgr stream velocity.  The direction of
the stream will be known better and better as further observations of
stars in the stream are made.

The count rate of the signal from the stream in the detector is also
modulated. For most of the recoil energy range, the phase of the count
rate from the stream is opposite to the phase of the characteristic
energy, i.e. for our standard case it peaks on June 28. For energies
near $E_c$, there is a small range of energy recoils in which the
count signal has the same phase as the characteristic energy and so
peaks in December.  The phase of the Sgr stream annual modulation is
different from that of the normal annual modulation of WIMPs in the
Galactic halo, which peaks on June 2. Contrary to the modulation of
the characteristic energy, the count rate of the stream signal depends
on the target nucleus, the WIMP mass, and the stream density, in
addition to the magnitude and direction of the Sgr stream velocity.

\bigskip
\subsection{ Directional detection of WIMPs}

A possibility of separating events due to WIMPs in the Sgr stream from those in
the local isothermal halo is to exploit detectors that can record the direction
of individual nuclear recoils (directional detection). It has been difficult to
build WIMP detectors sensitive to the direction of the nuclear recoils.
However, a recent promising development is the DRIFT detector \cite{DRIFT}. The
DRIFT detector consists of a negative ion time projection chamber, the gas in
the chamber serving both as WIMP target and as ionization medium for observing
the nuclear recoil tracks. The direction of the nuclear recoil is obtained from
the geometry and timing of the image of the recoil track on the chamber
end-plates.  A 1 m$^3$ prototype has been successfully tested, and a 10 m$^3$
detector is under consideration. The prototype contains $\sim 0.2$ kg of carbon
disulfide (CS$_2$) as active target.

The differential detection rate including the direction of the nuclear recoil
can be written using the results of \cite{radon} as
\begin{equation}
  \frac{ dR } { dE d\Omega_q} = \frac{ \sigma_0 F^2(q) }{ 4 \pi m \mu^2 }
 \,  \, \bigl[ \rho_h \radon{f}_h(v_{\rm min},\uvec{q}) + 
               \rho_{\rm str} \radon{f}_{\rm str}(v_{\rm min},\uvec{q}) \bigr],
\end{equation}
where $\uvec{q}$ is a unit vector in the direction of the nuclear recoil,
$d\Omega_q$ is an infinitesimal solid angle in the direction $\uvec{q}$,
$v_{\rm min}$ is defined above (Eq.\ref{eq:vmin}), and
\begin{equation}
  \radon{f}(p,\uvec{q}) = \int \delta ( p - \uvec{q} \cdot {\bf v}) 
  \, f({\bf v}) \, d^3 v ,
\end{equation}
is the 3-dimensional Radon transform of the velocity distribution function
$f({\bf v})$ (in our case, of $f_h({\bf v})$ and $f_{\rm str}({\bf v})$,
respectively).  

The Radon transforms of the velocity distributions we need have been computed
in \cite{radon}.
For the truncated Maxwellian distribution we assume for the WIMPs in the
isothermal halo, we have 
\begin{equation}
  \radon{f}_h(v_{\rm min},\uvec{q}) = 
  \frac{1}{N_{\rm esc}\pi^{1/2} \overline{v}_0}
  \left[ e^{-[v_{\rm min}+\uvec{\bf q}\cdot {\bf v}_\oplus(t)]^2/\overline{v}_0^2}
   - e^{-v_{\rm esc}^2/\overline{v}_0^2}
  \right] .
\end{equation}
For the WIMPs in the Sgr stream, assuming a Maxwellian velocity distribution
with velocity dispersion $\sigma_{\rm str}$, we have
\begin{equation}
  \radon{f}_{\rm str}(v_{\rm min},\uvec{q}) = 
  \frac{1}{\sqrt{2\pi\sigma_h^2/3}} 
  e^{-3[v_{\rm min}-\uvec{q}\cdot{\bf u}_{\rm str}]^2/2\sigma_h^2} ,
\end{equation}
where 
\begin{equation}
  {\bf u}_{\rm str} = {\bf v}_{\rm str} - {\bf v}_\oplus(t) .
\end{equation}
When the velocity dispersion is neglected, the Radon transform for the
stream reduces to
\begin{equation}
  \radon{f}_{\rm str}^{(0)}(v_{\rm min},\uvec{q}) = 
  \delta( v_{\rm min} - \uvec{q} \cdot {\bf u}_{\rm str}) 
\end{equation}
The latter case has a simple geometrical interpretation in the space
of vectors $v_{\rm min} \uvec{q}$. The Dirac delta function forces the
velocity $v_{\rm min}$ to change with the recoil direction $\uvec{q}$
according to $ v_{\rm min} = \uvec{q} \cdot {\bf u}_{\rm str}$. As
$\uvec{q}$ varies, the vectors $v_{\rm min} \uvec{q}$ describe the
surface of a sphere centered in $\frac{1}{2}{\bf u}_{\rm str}$ and
passing through the origin. The sphere has diameter $|{\bf u}_{\rm
  str}|$, and it is symmetric for rotations around ${\bf u}_{\rm
  str}$.

With a small velocity dispersion in the Sgr stream, the recoil vectors
$v_{\rm min} \uvec{q}$ due to WIMPs in the stream are still
approximately concentrated around the sphere just described.  On the
contrary, the recoils due to WIMPs in the isothermal halo are smoothly
distributed over a wide region of $v_{\rm min} \uvec{q}$ space. It is
this big difference in the recoil distributions from WIMPs in the
isothermal halo and in the Sgr stream that may allow their separation.

For detectors consisting of different nuclei, like e.g. CS$_2$ for
DRIFT, the total rate is a weighted average of the rates for the
individual nuclei with weights equal to the mass fraction of each
nucleus in the detector. In this case it is not possible to know 
with which type of nucleus the WIMP scattered in a given event. Hence it
is not possible to obtain the magnitude of the nuclear momentum
$q=\sqrt{2ME}$ or the velocity $v_{\rm min}=\sqrt{ME/(2\mu^2)}$, since
they depend on the nuclear mass $M$. In this case, it is convenient to
present the directional recoil spectrum $dR/dEd\Omega_q$ in terms of
the quantity
\begin{equation}
  {\bf E} = E \uvec{q} ,
\end{equation}
which is a vector with magnitude equal to the recoil energy and direction
pointing in the recoil direction. We call it the directional recoil energy.
The $(X,Y,Z)$ components of the directional recoil energy are the product of
the recoil energy and the direction cosines of the nuclear recoil
momentum. Since the volume element in the ${\bf E}$ space is $E^2 dE
d\Omega_q$, the relevant rate to plot is
\begin{equation}
  \frac{1}{E^2} \frac{ dR } { dE d\Omega_q} ,
\end{equation}
which is measured in counts per kg-day per keV$^3$.

\section{Results}

In Table II, we list the values of the characteristic energy $E_c$
of the step in the energy recoil spectrum due to WIMPs in the Sgr stream
appropriate to our reference case ($v_{str} = 300$ km/sec toward 
$(l,b)=(90^\circ,-76^\circ)$) for different target nuclei and WIMP masses.  As
discussed above, $E_c$ experiences an annual modulation, with a
maximum in June and a minimum in December. 

\begin{table}[tbp]
\begin{tabular}{|l||r|r||r|r||r|r|}
\hline
Target nucleus &
  \multicolumn{2}{c}{$m$=60 GeV} \vline &
    \multicolumn{2}{c}{$m$=100 GeV} \vline & 
      \multicolumn{2}{c}{$m$=500 GeV} \vline \\
\cline{2-7}
& 
  $E_c({\rm Dec 27})$ &$E_c({\rm Jun 28})$ &
    $E_c({\rm Dec 27})$ &$E_c({\rm Jun 28})$ &
      $E_c({\rm Dec 27})$ &$E_c({\rm Jun 28})$ \\
\hline
${}^{73}$Ge &
  42.3 & 35.1 &
    68.2 & 56.7 &
      149.7 & 124.3 \\
\hline
${}^{23}$Na &
  33.0 & 27.4 &
    41.2 & 34.3 &
      56.0 & 46.5 \\
\hline
${}^{127}$I &
  37.8 & 31.4 &
    70.2 & 58.3 &
      219.6 & 182.4 \\
\hline
${}^{40}$Ca &
  40.2 & 33.4 &
    56.1 & 46.6 &
      91.7 & 76.2 \\
\hline
${}^{183}$W &
  32.6 & 27.1 &
    65.9 & 54.7 &
      267.0 & 223.4 \\
\hline
${}^{16}$O &
  27.1 & 22.5 &
    32.1 & 26.6 &
      40.0 & 33.2 \\
\hline
${}^{131}$Xe &
  37.5 & 31.1 &
    70.0 & 58.2 &
      223.8 & 185.9 \\
\hline
${}^{27}$Al &
  31.6 & 29.4 &
    45.6 & 37.9 &
      64.8 & 53.8 \\
\hline
\end{tabular}
\caption{Characteristic recoil energies $E_c$ in keV 
from WIMPs in the Sagittarius stream.  Note that the value of
$E_c$ has an annual modulation with a peak on December 27 and 
a minimum on June 28. Numbers have been obtained for our reference
case, with a stream velocity of 300 km/sec toward $(l,b)=(90^\circ,-76^\circ)$.} 
\end{table}

The annual modulation of the characteristic energy $E_c$ can in
principle be used to identify the signal due to WIMPs in the Sgr
stream.  In particular, the annual modulation is a useful tool to
differentiate between an increase in the count rate due to WIMPs in
the Sgr stream at energies below $E_c$, and an increase due to
modifications of the velocity distribution of Galactic halo WIMPs.
These possibilities are difficult to distinguish by simply examining
the shape of the recoil spectrum because of the finite energy
resolution and the binning of data in recoil energy in a real
experiment. For this tool to be effective, the energy resolution must
of course be better than the difference between the maximum and
minimum values of $E_c$. As discussed in Section 3.4 above, this
difference corresponds to an energy resolution of 2-18\% depending on
the exact velocity of the Sgr stream. For our reference case of
$v_{str} = 300$ km/sec in the direction $(l,b)=(90^\circ,-76^\circ)$,
the value of $E_c$ is modulated by 9\% over the course of the year.
As long as the uncertainty in energy due to the energy resolution is
smaller than the energy shift in $E_c$, one should be able to detect
its annual modulation.

We note that the values of $E_c$ are indeed in the range accessible to
current and upcoming dark matter experiments.  For example, the NaI
detector of the DAMA experiment \cite{DAMA} has an energy threshold of
2 keV electron equivalent, which corresponds to 22 keV recoil energy
for iodine; since iodine is much heavier than sodium, WIMP
interactions with iodine dominate the count rate, and the
conversion factor between keVee and keV is a factor of 11 for iodine.
Hence for WIMP masses heavier than 50 GeV, the step lies above the
DAMA threshold energy, and DAMA can in principle detect it. The
question is whether or not DAMA can in fact identify an increase in
count rate with a signal from the Sgr stream. 

\begin{figure}
\includegraphics[width=\textwidth]{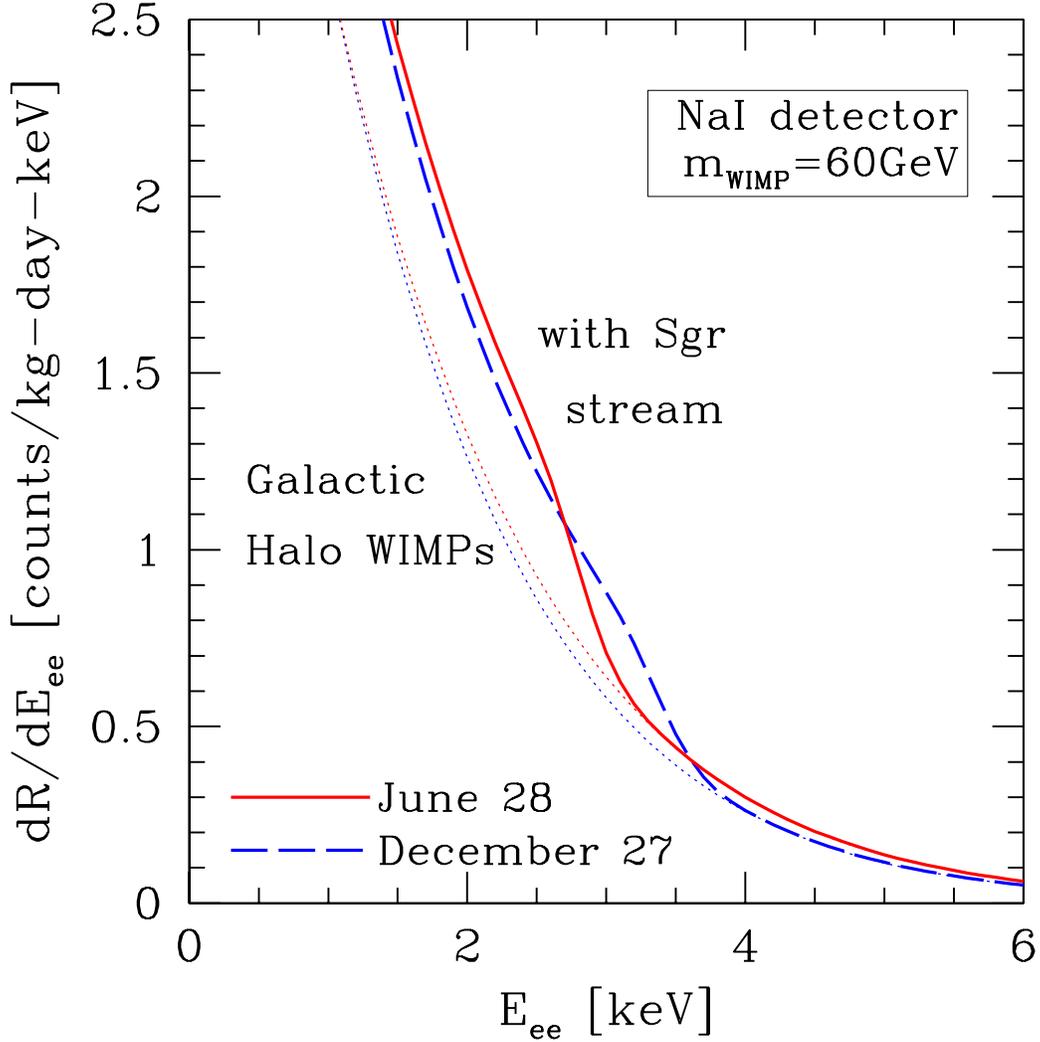}
\caption{
  Count rate of 60 GeV WIMPs in a NaI detector such as DAMA as a
  function of recoil energy.  The dotted lines (towards the left)
  indicate the count rate due to Galactic halo WIMPs alone for an
  isothermal halo.  The solid and dashed lines indicate the step in
  the count rate that arises if we include the WIMPs in the Sgr stream
  for $v_{str} = 300$ km/sec in the direction
  $(l,b)=(90^\circ,-76^\circ)$ with a stream velocity dispersion of 20
  km/sec.  The plot assumes that the Sgr stream contributes an
  additional 20\% of the local Galactic halo density.  The solid and
  dashed lines are for June 28 and December 27 respectively, the dates
  at which the annual modulation of the stream is maximized and
  minimized.  }

\end{figure}

In Figure 1, we have plotted the count rate of 60 GeV WIMPs in a
NaI detector such as DAMA as a function of
recoil energy.  The dotted lines (towards the left) indicate the count
rate due to Galactic halo WIMPs alone for an isothermal halo.  The
solid and dashed lines indicate the count rate including the WIMPs in
the Sgr stream for our reference case of $v_{str} = 300$ km/sec in the
direction $(l,b)=(90^\circ,-76^\circ)$ with a stream velocity
dispersion of 20 km/sec.  In this plot, we have taken the Sgr stream
to contribute an additional 20\% to the local Halo density of our Galaxy.  As
discussed earlier, the best estimate of the contribution varies from
0.3\% to 23\%.  The solid and dashed lines are for June 28 and December
27 respectively, the dates at which the stream annual modulation is
maximized and minimized.  As noted above, the annual modulation of the
value of $E_c$ for this reference case is a 9.2\% effect, and peaks at
a different date than the standard annual modulation in the Galactic
halo (which causes the count rate to peak on June 2).  

The DAMA experiment has an enormous exposure of 107731 kg-days.  For a
60 GeV WIMP mass and a maximal Sgr stream that is 20\% of the local
halo density, we compute the following number of events expected in 1
keV bins (2-3 keVee, 3-4 keVee, 4-5 keVee, 5-6 keVee), averaged over a
year: (136744, 51814, 20873, and 9367) with the Sgr stream, and
(98370, 45745, 20717, and 9214) without the Sgr stream.\footnote{For
  these stream parameters, iodine does not provide a signal from the
  stream beyond 5 keVee, while sodium gives rise to a long but weak
  tail of stream events extending up to 21 keVee. This tail, however,
  contributes less than one event per keVee in 107731 kg-days beyond
  11 keVee. The long tail is due to the very different quenching
  factors of iodine and sodium: 0.09 and 0.3, respectively.} The
location of the step is roughly at 3 keVee.  For a 60 GeV WIMP mass
and a stream that is 20\% of the local halo, in the 2-3 keVee bin, the
difference in number of counts with and without the stream is 38374,
which is 104 times the square root of the number of counts with the
stream. Hence we find that, in the 2-3 keVee bin, the presence of the
stream is detectable at the 104$\sigma$ level. In the 3-4 keVee bin,
we find that the stream is detectable at the $27\sigma$ level. These
results have assumed that the density of WIMPs in the Sgr stream is
20\% of the local halo.  If, as is more reasonable, the density of
WIMPs in the stream is instead 4\% of the local halo, the stream is
detectable at the $24\sigma$ and $6\sigma$ level in the 2-3 and 3-4
keVee bins respectively.  In more detail, if the stream density is 4\%
of the local halo, the number of expected events in 1 keV bins from 2
to 6 keVee, averaged over a year with an exposure of 107731 kg-days,
is (106045, 46959, 20748, and 9245). The expected count rates for
other stream densities can easily be obtained by scaling the stream
contribution in the numbers quoted above, for the contribution of the
stream scales proportionally with the ratio of stream density to local
halo density.

%Since the numbers for the data are
%not published, it is difficult for us to determine whether or not the
%stream is in the data.  We encourage the DAMA experimenters to look
%for the step due to the Sgr stream.  

%The DAMA experiment finds that the count rate in their data peaks on
%May 21 $\pm$ 22 days (1$\sigma$ error bars \cite{DAMA}), 
%i.e. a day between April 29 and June 12
%(2 $\sigma$ errors include the range April 7 to July 4). 
%The error band includes June 2,  
%a date which is predicted by an isothermal halo model.  
We
note that, in the presence of a stream, the date of the maximum count
rate changes. For the reference case of $v_{str} = 300$ km/sec in the
direction $(l,b)=(90^\circ,-76^\circ)$ the count rate due to the WIMPs
in the stream alone peaks on June 28 and December 27 at different
values of recoil energy.  Depending on the contribution of stream
WIMPs relative to the local halo WIMPs, the overall count rate may
peak between these dates and the usual June 2.  For a Sgr WIMP density
that is 20\% of the local halo, the overall count rate peaks on June
28 and December 27 for different values of the recoil energy.  For
lower values of the stream density, the date of peak count rate
changes and eventually returns to the same value as without a stream.
A further investigation of this sensitivity is in progress.  
%DAMA may
%in fact be used to constrain the Sgr dwarf stream parameters.
%rule out a high density Sgr stream for our
%reference case of $v_{str} = 300$ km/sec in the direction
%$(l,b)=(90^\circ,-76^\circ)$.

%In addition, DAMA may be able to see the annual modulation of the
%step, which (as discussed above) is important to proving that they
%have indeed seen the Sgr stream.  With their current energy resolution
%of 7.5\%, they may be able to make this identification in some cases
%but not in others.  For the reference case of $v_{str} = 300$ km/sec
%in the direction $(l,b)=(90^\circ,-76^\circ)$, the energy resolution
%is better than the difference in $E_c$ due to annual modulation.  In
%particular, DAMA should be able to use the modulation in $E_c$ in the
%case of Table II, which corresponds to the reference case for the
%stream and a WIMP mass of 60 GeV.  Thus the DAMA experiment may
%already have the Sgr modulation in their current data.  Finding the
%Sgr modulation may be an important step in interpreting the observed
%annual modulation in DAMA in terms of WIMPs.

%The existence of the Sgr stream in the DAMA data may shift the best
%fit WIMP mass. If a stream is indeed found to be present in the data,
%then the maximum likelihood estimates would need to be redone
%to ascertain the new best fit WIMP mass.  Indeed it is possible
%that the discrepancy between DAMA and Edelweiss data may be resolved
%if the stream is correctly included.  A further study of this
%proposed resolution is in progress.

The sapphire (Al$_2$O$_3$) and calcium tungstate (CaWO$_4$) detectors
of CRESST \cite{CRESST} have a better energy resolution. Its smaller
detector mass requires a longer exposure.  The energy threshold is
low, $\sim 1$ keV.

The ${}^{73}$Ge detector of the CDMS experiment \cite{CDMS} has a
low energy threshold at 10 keV and a good energy resolution, about 0.5
keV at threshold and about 1.5 keV at 100 keV. As can be seen from
Table II and Figure 1, which shows the energy recoil spectrum in
${}^{73}$Ge, the characteristic energy of the reference model varies
from 25-120 keV, easily above the energy threshold of the experiment.
In addition, the CDMS energy resolution in this range varies between
2\% and 4\%, so the CDMS experiment has the required energy resolution
to be sensitive to the annual modulation of the Sgr stream for most of
the range of expected velocities.  The EDELWEISS experiment
\cite{EDELWEISS}, also with a ${}^{73}$Ge detector, has an energy
threshold of 20 keV.
% and energy resolution of .., which for our
% purposes are comparable to those of CDMS.

CDMS-II at Soudan is expected to have an exposure of 2500 kg-day
\cite{CryoArray}.  Using the same parameters as in Figure 2, in
particular a WIMP mass $m=60$~GeV, we find the following number of
events in 10 keV bins (10-20 keV, 20-30 keV, 30-40 keV, 40-50 keV,
50-60 keV), averaged over a year: (3083, 2138, 1386, 666, 383) with
the Sgr stream and (2473, 1592, 1006, 623, 383) without the Sgr
stream. The last bin has the same number of counts with or without the
stream because it is well above the characteristic energy of the step
(in this case, $E_c$ varies between 35.1 and 42.3 keV).  In the first
bin, the difference in number of counts with and without the stream is
610, which is 11 times the square root of the number of counts with
the stream. Hence we find that, in the first bin, the presence of the
stream is detectable at the 11$\sigma$ level. In the second and third
bins we find that the stream is detectable at the $12\sigma$ and
$10\sigma$ level, respectively. This assumed that the density of WIMPs
in the Sgr stream is 20\% of the local halo density. With 4\% density,
instead, the stream should be detectable in the same bins at the
$2.4\sigma$, $2.7\sigma$, and $2.3\sigma$ levels, respectively.  With
the proposed CryoArray detector, which will have 500000 kg-day of data
\cite{CryoArray}, the Sgr stream should be easy to identify.

\begin{figure}
\includegraphics[width=\textwidth]{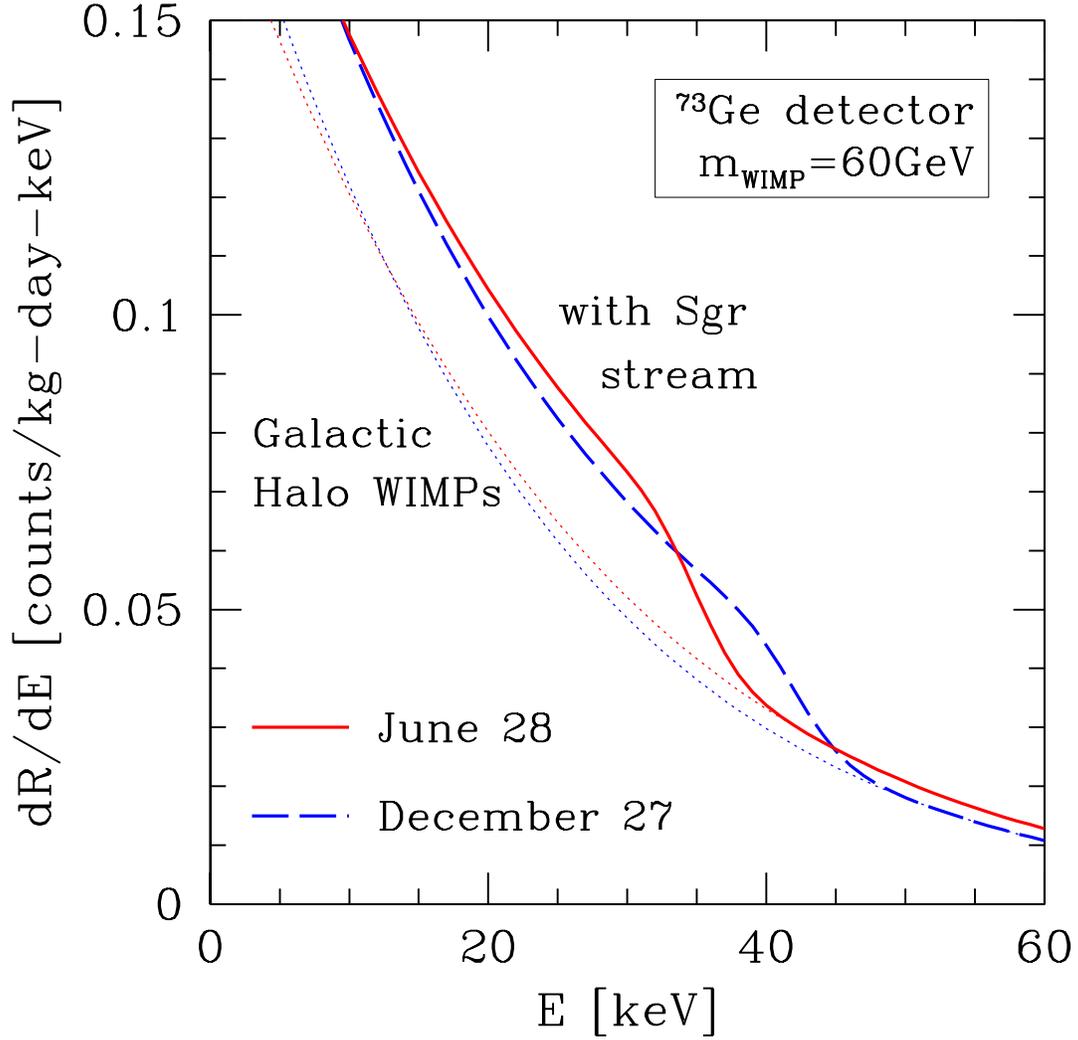}
\caption{
  Count rate of 60 GeV WIMPs in a ${}^{73}$Ge detector as a function
  of recoil energy.  The dotted lines (towards the left) indicate the
  count rate due to Galactic halo WIMPs alone for an isothermal halo.
  The solid and dashed lines indicate the step in the count rate that
  arises if we include the WIMPs in the Sgr stream.  The parameters of
  the curves are the same as described in Figure 1. }
\end{figure}

In Figure 2, we have plotted the count rate of 60 GeV WIMPs in a
${}^{73}$Ge detector (such as CDMS or Edelweiss) as a function of
recoil energy.  The parameters of the curves are the same as in Figure
1.  The dotted lines (towards the left) indicate the count rate due to
Galactic halo WIMPs alone for an isothermal halo.  The solid and
dashed lines indicate the count rate including the WIMPs in the Sgr
stream for our reference case.  The characteristic energy of the step
is 35.1 keV in June and 42.3 keV in December.
%Due to the fact that the total flux
%of WIMPs in the stream is constant during the year, the area under
%the solid and dashed curves must be roughly the same; hence the step due to
%the higher count rates at low energy in June must end at a lower $E_c$
%than in December.

Upcoming detectors XENON \cite{XENON} and ZEPLIN-II \cite{ZEPLIN}
have low energy thresholds of 2-4 keV. With 1 ton detectors made of
liquid xenon planned, these experiments should see enormous count
rates and be able to distinguish the Sgr stream.

\begin{figure}
\includegraphics[width=\textwidth]{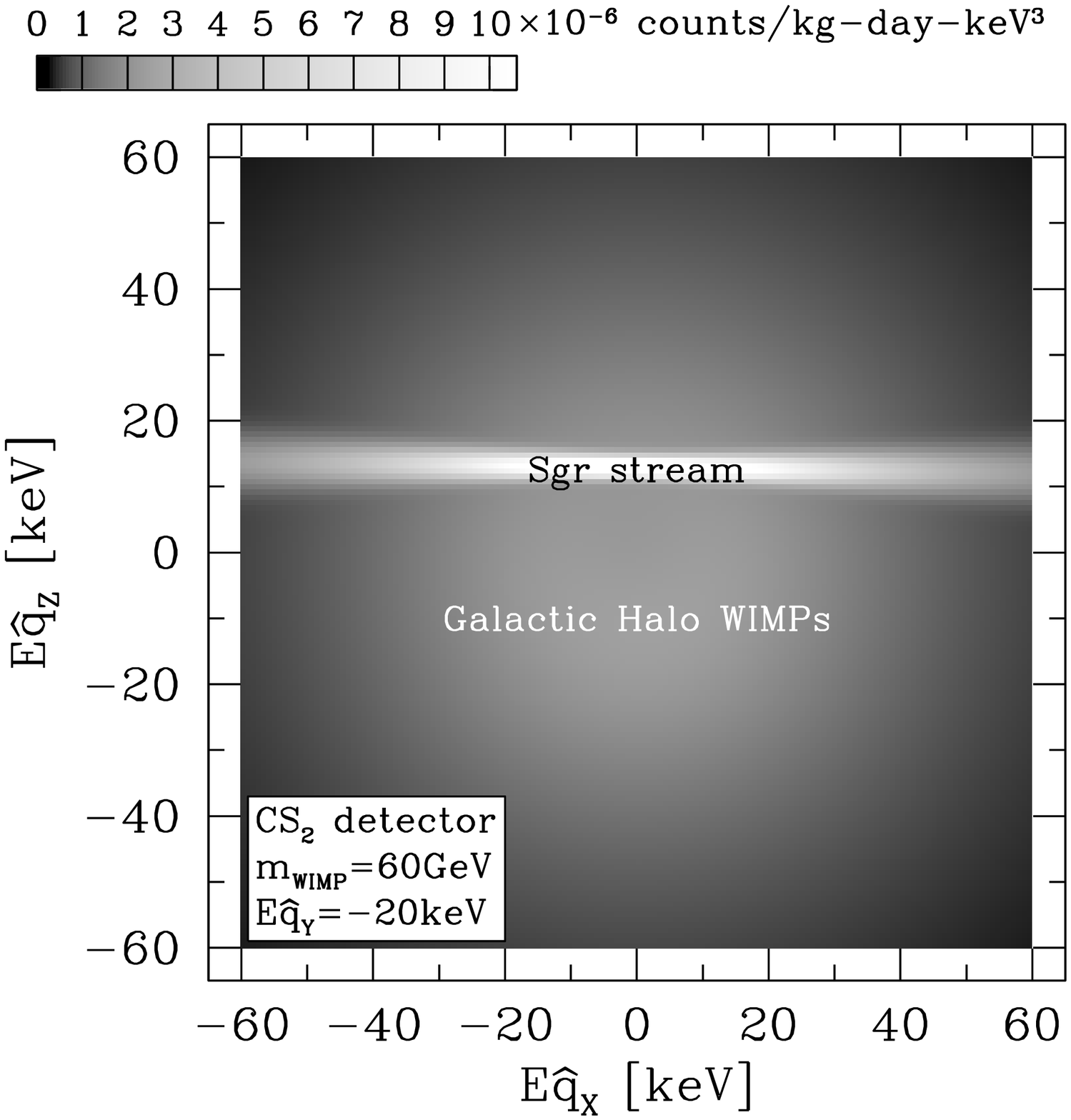}
\caption{
  Count rate of 60 GeV WIMPs in a CS$_2$ detector (DRIFT) as a
  function of recoil energy and direction of the nuclear recoil. The
  figure shows the count rate in a 2-dimensional slice of the
  3-dimensional recoil space.  The chosen slice is perpendicular to
  the direction of Galactic rotation and defined by a recoil energy of
  20 keV in that direction. The horizontal axis represents recoils in
  the direction of the Galactic center (left) and Galactic anticenter
  (right); the vertical axis represents recoils in the direction of
  the North Galactic Pole (upward) and South Galactic Pole (downward).
  The gray scale indicates the count rate per kilogram of detector per
  day and per unit cell in the 3-dimensional energy space. Lighter
  regions correspond to higher count rates. The white band on the
  upper part is the location of nuclear recoil due to WIMPs in the Sgr
  stream. The fuzzy gray cloud at the center contains recoils due to
  WIMPs in the local isothermal Galactic halo. The two WIMP
  populations can in principle be easily separated, given a sufficient
  exposure.}
\end{figure}

\begin{figure}
\includegraphics[width=\textwidth]{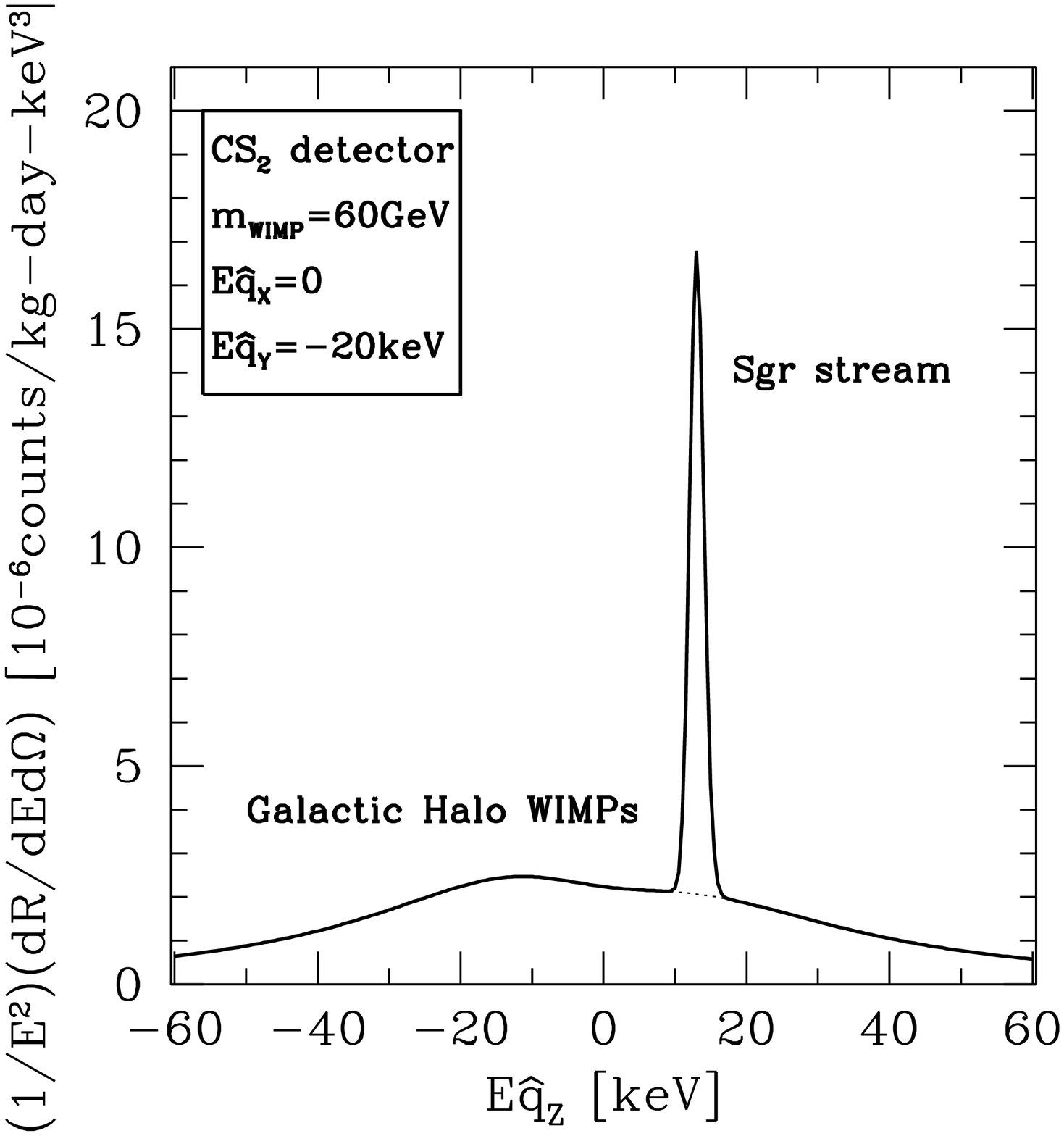}
\caption{
  Section of figure 3 with $E\uvec{q}_X=0$.}
\end{figure}

In Figure 3 we illustrate the power of directional detectors. The
figure shows the count rate of 60 GeV WIMPs in a CS$_2$ detector
(DRIFT) as a function of recoil energy and direction of the nuclear
recoil.  As above, we have assumed that the Sgr stream impinges on the
Galactic plane with a speed of 300 km/s in the direction
$(l,b)=(90^\circ,-76^\circ)$.  We plot the rate at the maximum of the
stream modulation on December 27.  The figure shows the count rate in
a 2-dimensional slice of the 3-dimensional recoil space.  The chosen
slice is perpendicular to the direction of Galactic rotation and
defined by a recoil energy of 20 keV in that direction.  The
horizontal axis represents recoils in the direction of the Galactic
center (left) and Galactic anticenter (right); the vertical axis
represents recoils in the direction of the North Galactic Pole
(upward) and South Galactic Pole (downward). The gray scale indicates
the count rate per kilogram of detector per day and per unit cell in
the 3-dimensional energy space. Lighter regions correspond to higher
count rates. The white band on the upper part is the location of
nuclear recoil due to WIMPs in the Sgr stream.  The fuzzy gray cloud
at the center contains recoils due to WIMPs in the local isothermal
halo. The two WIMP populations can in principle be easily separated,
given a sufficient exposure.  Figure 4 shows a section of Figure 3
(as defined in the captions) and illustrates that the Sgr
stream clearly stands out above the Galactic halo WIMPs.

We can estimate the exposure needed to identify the Sgr stream by
considering binning directional data into cells in directional energy
space of size 10 keV.  The count rate per bin is approximately given
by the rate in figures 3 and 4 multiplied by the volume of the cell
($10^3$ keV$^3$). The count rate for the Sgr stream results in $\sim
10^{-2}$ counts/kg-day/cell, and the count rate for the isothermal
component is $\sim 3 \times 10^{-3}$ counts/kg-day/cell. Using a
$\sqrt{N}$ argument, we see that the Sgr stream may be distinguished
at the 3$\sigma$ level with an exposure of $\sim 10^4$ kg-days. For
DRIFT, this corresponds to exposing a 30 m$^3$ of CS$_2$ for 5 yr.
This is in the upper range of the current plans for DRIFT-2, whose
volume is expected to be in the range 10-30 m$^3$.

Therefore DRIFT has the capability of identifying WIMPs in the Sgr stream.

\section{Conclusions}

Recent observations of the Sagittarius dwarf spheroidal galaxy (Sgr)
indicate the existence of tidal streams of stars that pass through the
solar neighborhood.  If the mass-to-light ratio in the streams is at
least comparable to that of the main body of the Sgr galaxy, then one
can detect the existence of dark matter in the streams.  Under the
assumption that the stream passes through the solar position, the dark
matter stream plausibly has a density $\sim $(0.3-23)\% of the local density of
our Galactic Halo.  If dark matter consists of WIMPs, the extra
contribution from the stream gives rise to a step-like feature in the
energy recoil spectrum in direct dark matter detection, i.e., an
increase in the rate in detectors at energy recoils less than some
critical value $E_c$ that depends on the target nucleus and on the
time of year. The location of the step experiences an annual
modulation that will be useful in identifying the existence of the
stream.  The count rate in the detector is also modulated annually,
with a maximum on June 28 and minimum on December 27 except near the
characteristic energy where the phase is opposite.

%For WIMP masses heavier than 50 GeV, the step lies above the threshold
%energy of the DAMA experiment, and DAMA may have it in their current
%data.  In the current DAMA data, for a 60 GeV WIMP mass and a very
%reasonable stream density that is 4\% of the local halo, the stream is
%detectable at the 24$\sigma$ level in the 2-3 keV electron equivalent
%energy bin.  For the most favorable stream density (20\% of the local
%halo), the stream is detectable at the 104$\sigma$ level in the 2-3
%keV electron equivalent energy bin; however, in this case the count
%rate peaks on a different date than that of an isothermal halo and for 
%some combinations of stream densities and velocities may
%be in disagreement with the current DAMA data.  A study of the
%dependence of the date of the peak count rate on the amount of
%contribution due to a stream is in progress.  We encourage the
%experimenters to look for the stream in their data.  If the existence
%of the Sgr stream were confirmed in the data, then a new best fit WIMP
%mass would have to be obtained.  It is conceivable that the stream is
%responsible for an apparent discrepancy between DAMA and other
%experiments such as CDMS, EDELWEISS, and ZEPLIN-I.  In addition, the DAMA experiment can
%check for the annual modulation of the step due to the Sgr stream dark
%matter.  Finding the Sgr modulation may be an important step in
%interpreting the observed annual modulation in DAMA in terms of WIMPs.

In the CDMS experiment, for our best estimate of stream velocity (300
km/sec) and direction [$(l,b)=(90^\circ,-76^\circ)$], the location of
the step oscillates yearly between 25 and 35 keV, with a phase
opposite to that of the count rate.  With two kg of Germanium, the
stream should be detectable at three $\sigma$'s in four years of data
with 10 keV energy bins.  Planned large detectors like XENON,
CryoArray and the directional detector DRIFT may be able to identify
the Sgr stream.

%\acknowledgments

K. F. acknowledges support from the DOE via a grant in the Physics
Dept. at the Univ. of Michigan. K.F.  and P.G. thank the Michigan
Center for Theoretical Physics for support.  H. N. acknowledges
support from Research Corporation and the National Science Foundation
(AST-0307571).
%The Sgr stream peaks in June.

\end{document}